# Optical structure and polarization of the western hot spot of Pictor A[1]

R. C. Thomson

Institute of Astronomy, Madingley Road, Cambridge CB3 0HA, UK.

P. Crane

European Southern Observatory, Karl Schwarzschild Strasse 2, D-8046 Garching, Germany.

C. D. Mackay

Institute of Astronomy, Madingley Road, Cambridge CB3 0HA, UK.

astro-ph/9505122   26 May 1995

---

# ABSTRACT

HST images of the western hot spot of the nearby FRII radio galaxy Pictor A resolve the hot spot into an amorphous region of highly polarized wisps. The high polarization ($\gtrsim 50$ %) indicates that the optical emission is synchrotron radiation and that the magnetic field structure has been well resolved. The magnetic field is oriented nearly perpendicular to the jet axis and apparently very uniform. We interpret these observations in terms of a compressed magnetic field at the working surface where the jet impinges on the intergalactic medium some 100 kpc from the nucleus of the host galaxy.

*Subject headings:* Polarization — radiation mechanisms: nonthermal — galaxies: active — galaxies: individual: Pictor A — galaxies: jets.

1. Introduction

Pictor A is the nearest broad line radio galaxy with a Fanaroff and Riley class II (FRII) radio morphology. At a redshift of 0.035 it lies at a distance of some 140 Mpc ($H_0$=75 km s$^{-1}$ Mpc$^{-1}$) where 1 arcsec corresponds to 700 pc. The radio lobes extend over 100 kpc on either side of the host galaxy at a position angle of 102°. Plate 1a shows the optical field around Pictor A (including the host galaxy and both the western and the eastern hot spots) with contours of the radio emission superimposed. Recently, Simkin et al. (1993, 1994) obtained a high resolution (1 arcsec) radio map of the central source using the Australia Telescope Compact Array which reveals a kiloparsec scale radio jet that points towards the western hotspot.

Plate 1b shows a close up of the radio emission from the western hot spot of Pictor A, and Plate 1c shows the optical emission from the same region. Although some extended low surface brightness emission is seen, the hot spot itself is unresolved at optical wavelengths and appears to sit outside the boundary of the western radio lobe (Meisenheimer et al. 1989). Röser & Meisenheimer (1987) found that the optical emission from the hot spot has a featureless blue spectrum with a flux density consistent with the extrapolation of the non-thermal radio spectrum. They also detected significant linear polarization, indicating that the optical emission from the hotspot is most likely to be synchrotron radiation.

Since the structure of the western hot spot of Pictor A is unresolved from the ground at optical wavelengths, it is a prime target for high resolution imaging using the Faint Object Camera (FOC) on the Hubble Space Telescope (HST). Full advantage has been taken of the FOC facilities by including linear polarization measurements to determine the magnetic field structure in the hot spot. The results show that the hot spot is composed of wisps that are elongated nearly perpendicular to the jet axis. The fractional polarization is extremely high ($\gtrsim 50$ %) and reveals a well resolved magnetic field structure which is aligned along the direction of elongation of the wisps, and nearly perpendicular to the jet axis.

## 2. Observations and Data Reduction

Three linearly polarized images of the western hot spot of Pictor A were obtained on 5 February 1994 using the FOC f/96 camera with the B band filter (F430W) and the polarizers POL0, POL60 and POL120 consecutively. The exposure times were 991, 1076, and 895 secs respectively. The typical background level in each image is 0.8 counts pixel$^{-1}$ with a peak source count of 13.6 counts pixel$^{-1}$.

These data were some of the first to be obtained using the FOC after the refurbishment mission. The post-COSTAR optics increase the focal ratio of the f/96 camera to f/151, although it is still referred to as the f/96 camera. For the normal $512 \times 512$ format, as used here, the pixel size is $0.014 \times 0.014$ arcsec$^2$ and the field of view is $7.1 \times 7.1$ arcsec$^2$. The instrument resolution is 0.035 arcsec (FWHM), as measured using a point source with the same F430W filter. The inclusion of the polarizers is known to degrade the resolution slightly, and we estimate the effective resolution of these data to be 0.04 arcsec.

The images were flat fielded and geometrically corrected using the default calibration data, as distributed to us by the Space Telescope Science Institute, and the mean background level subtracted from each polarized image. Before combining the polarized images, they were first aligned by taking into account the small but significant offsets introduced when the polarizers are inserted in the optical path (Hodge 1993). When used in conjunction with the filter F430W, the relative offsets are ($\Delta x, \Delta y$; measured in pixels): 0.0,0.0; -3.5,-1.7; 0.0,0.8 for the polarizers POL0, POL60 and POL120 respectively. It is also known that POL60 has a low transmittance below 3000 Å, but this is not a problem at this longer wavelength where the transmittances match to better than 1% (Hodge 1992).

The aligned images were scaled to the same exposure time (1000 s) before combining to produce the total intensity image and polarization map. The procedure for doing this is described by Thomson et al. (1995). For the polarization map, the images were binned over $8 \times 8$ pixels to increase the S/N. The resulting polarization map has an effective resolution of just over 0.1 arcsec. We estimate the relative error in the fractional polarizations to be $\lesssim 6\%$ with a corresponding position angle error of $\lesssim 3^0$.

## 3. Results

Plate 2 shows the total intensity image after binning over $2 \times 2$ pixels. This was done to increase the image S/N. The hot spot is well resolved by the HST with a diameter of 1.4 arcsec, although the brightest region is only 0.3 arcsec across (FWHM). The hot spot appears to be composed of wisps elongated nearly perpendicular to the jet axis. There is some evidence of more extended low surface brightness emission, but the FOC images have insufficient S/N to trace such faint emission.

Fig. 1 shows the polarization map superimposed on a contour plot of the total intensity image. The high fractional polarization ($\gtrsim 50$ %) observed throughout the hot spot implies that the radiation mechanism is most likely synchrotron radiation with the magnetic field well aligned along the direction of the elongated wisps and nearly perpendicular to the jet axis. Any unresolved random field component must be smaller than the resolved field component.

## 4. Discussion

Meisenheimer et al. (1989) found a power law spectrum with a spectral index $\alpha=0.89$ ($S_\nu \propto \nu^{-\alpha}$) for the radio through optical emission from the western hot spot of Pictor A.

This implies a maximum observable polarization of 74% for a completely uniform magnetic field. The observed high polarizations of $\gtrsim 50$ % imply that the magnetic field structure is almost completely resolved and appears to be aligned parallel to the direction of elongation of the wisps in the hot spot. Assuming that the resolved field component is completely uniform ($B_{uni}$) and the unresolved component is completely random ($B_{ran}$), then the observed fractional polarization implies that $B_{ran} \lesssim B_{uni}$.

The discovery of a sub-arcsecond optical jet (Simkin et al. 1992), and an arcsecond radio jet (Simkin et al. 1993) located between the inner optical jet and the western hot spot of Pictor A are clear indications that an otherwise invisible jet powers the optical and radio synchrotron emission from the western hot spot. As with other FRII radio sources, the actual jets are intrinsically very faint and only become visible when observed with sufficient sensitivity and angular resolution. In this case, the jet remains undetected outside the confines of the host galaxy nucleus and the western hot spot. We think it is most likely that the western hot spot of Pictor A represents the working surface where the jet hits the intergalactic medium.

Blandford & Rees (1974) first proposed that hot spots are the working surface where a jet impinges on the intergalactic medium. The structure of such a working surface comprises three main features: a strong shock in the jet itself, a bow shock in the surrounding medium, and a contact discontinuity where exhaust gases from the jet mix with the post-shock intergalactic medium. Numerical simulations confirm this general structure and the results have been compared with observations (eg, Smith et al. 1985). The structure and polarization of simulated hot spots depends crucially on viewing angle (Clarke, Norman & Burns 1989). The sharp edged appearance of the brightest wisp in the western hot spot of Pictor A together with the striking alignment of the magnetic polarization vectors perpendicular to the jet axis strongly suggest an almost edge-on viewing angle (ie, the jet must lie close to the plane of the sky). This is consistent with the observation that the western hot spot of Pictor A actually protrudes from the radio lobe (Meisenheimer et al. 1989), implying that it must be seen nearly edge-on. We conclude that the jets of Pictor A lie close to the plane of the sky.

The optical polarization of the western hot spot of Pictor A is similar to that of the eastern hot spot in M87 as measured by Sparks et al. (1992) with the position angle of the magnetic vector nearly perpendicular to the jet axis. In addition, the detailed radio map of M87 (Biretta 1993) shows wispy features in the eastern hot spot, similar to the western hotspot of Pictor A. High resolution polarization maps of the brightest knot (Knot A) in the visible jet on the western side of M87 also reveal magnetic field vectors nearly perpendicular to the jet axis (Owen, Hardee & Cornwell 1989; Thomson et al. 1995). Unlike the eastern hotspot of M87 and the western hot spot of Pictor A, however, Knot A does not occur at the end of the jet in M87, but appears to be a strong shock that slows the jet down but does not completely disrupt it (Biretta 1993). A high resolution optical polarization map of the jet of the quasar 3C 273 (Thomson, Mackay & Wright 1993) shows that magnetic field vector of the outermost optical knot is not perpendicular to the jet axis, but inclined at about $40^0$. Unlike the western hotspot of Pictor A, the radio emission from the jet of 3C 273 extends 1 arcsec (ie, 2.5 kpc) further from the quasar than does the optical emission, although the outermost radio hot spot is polarized perpendicular to the jet axis.

Our FOC observations have shown that the western hot spot of Pictor A is composed of wisps elongated in a direction nearly perpendicular to the jet axis. The high polarization shows that the magnetic field is well resolved and aligned with the wisps, nearly perpendicular to the jet axis. The results are consistent with the standard model in which the hot spots of FRII sources (like Pictor A) emit pure synchrotron radiation, powered by invisible jets which originate in the nucleus of the host galaxy. Accordingly, both the

– 8 –

optical and radio emission from these hot spots is generated by relativistic particles that are accelerated in the terminal shock which forms at the working surface between the jet and the surrounding medium.

Further observational work is currently hampered by the unavailablity of high resolution radio polarization data. When such data become available, they could be combined with our FOC data to produce high resolution spectral index, depolarization and rotation measure maps which will reveal the physical state of the hot spot and its environment.

We are grateful to Richard Perley for making available the 8.4 GHz VLA radio map (Plate 1b). We thank Robert Laing for useful discussions, and Richard Sword for help in preparing the figures. P.C. acknowledges support through NASA.

– 10 –

# Figure Captions

**Plate 1.** **a)**. Field around the host galaxy of Pictor A (reproduced from the digitized Sky Survey distributed by STScI), with the 1.4 GHz radio emission (contours) superimposed (radio map reproduced with permission from *Hot Spots in Extragalactic Radio Sources*, p. 100. Copyright ©1989 Springer-Verlag.) The location of the western hot spot is indicated by the box. **b)**. Close up of the 8.4 GHz radio emission from the western hot spot of Pictor A (reproduced with permission from R. Perley). The field shown is the boxed area in a), and the resolution is $0.8 \times 0.2$ arcsec$^2$. **c)**. Optical (R band) CCD image of the hot spot taken in 1 arcsec seeing (reproduced with permission from *Hot Spots in Extragalactic Radio Sources*, p. 100. Copyright ©1989 Springer-Verlag.) The field is the same as b).

**Plate 2.** B band (F430W) total intensity image of the western hotspot of Pictor A. The resolution is 0.04 arcsec (FWHM). This image was formed by adding the three linearly polarized images after alignment and weighting by the relative exposure times as described in the text.

**Figure 1.** Polarization map of the optical emission from the western hotspot of Pictor A superimposed on a contour plot of the total intensity image. The polarization map has a resolution of 0.1 arcsec and shows the orientation of the *magnetic* field vectors. The map was formed by combining the three linearly polarized images, as described in the text. The direction of the host galaxy nucleus is shown and this defines the jet axis.